\documentclass[journal, twoside]{IEEEtran}

\usepackage{amsmath}
\usepackage{dsfont}
\usepackage{amssymb,latexsym}
\usepackage{amsfonts}
\usepackage{amscd}
\usepackage{algorithm}
\usepackage{algorithmic}
\usepackage{multirow}
\usepackage{array}
\usepackage{changepage}
\usepackage{times}
\usepackage{float}
\usepackage{graphics,subfigure}
\usepackage{graphbox}
\usepackage[latin1]{inputenc}
\usepackage{graphicx}
\usepackage{color}
\usepackage{relsize,balance}
\usepackage{cite}
\usepackage{stfloats}
\usepackage{epstopdf}
\usepackage{setspace}

\newcommand{\bs}{\boldsymbol}

\newcommand{\mI}{{\bs I}}

\newcommand{\vb}{{\bs b}}

\newcommand{\vw}{{\bs w}}
\newcommand{\vx}{{\bs x}}

\newcommand{\vz}{{\bs z}}

\newcommand{\valpha}{{\bs \alpha}}
\newcommand{\vdelta}{\bs{\delta}}
\newcommand{\vkappa}{{\bs \kappa}}
\newcommand{\vomega}{{\bs \omega}}

\def\D{\ensuremath{\mathcal{D}}}
\def\E{\ensuremath{\mathbb{E}}}
\def\HH{\ensuremath{\mathcal{H}}}
\def\L{\ensuremath{\mathcal{L}}}
\def\R{\ensuremath{\mathbb{R}}}
\def\N{\ensuremath{\mathcal{N}}}
\def\U{\ensuremath{\mathcal{U}}}

\begin{document}

%
\title{Adaptive Random Fourier Features Kernel LMS}
%
%
%

\author{Wei~Gao,~\IEEEmembership{Member,~IEEE},
        Jie~Chen,~\IEEEmembership{Senior Member,~IEEE},
        C\'edric~Richard~\IEEEmembership{Senior Member,~IEEE},
        Wentao~Shi,~\IEEEmembership{Member,~IEEE}, and
         Qunfei~Zhang,~\IEEEmembership{Member,~IEEE}\\

\thanks{Manuscript received MM DD, 2022; revised MM DD, 2022.}
\thanks{This work was supported in part by the National NSFC under Grants 62171205 and 62171380.}
\thanks{Wei Gao is with the School of Computer Science and Telecommunication Engineering, Jiangsu University, Zhenjiang 212013, China (email: wei\_gao@ujs.edu.cn).}
\thanks{Jie Chen, Wentao Shi, and Qunfei Zhang are with the School of Marine Science and Technology, Northwestern Polytechnical University, Xi'an 710072, China (email: dr.jie.chen@ieee.org; swt@nwpu.edu.cn; zhangqf@nwpu.edu.cn).}
\thanks{C\'edric Richard is with the CNRS, OCA, Universit\'e C\^ote d'Azur, 06108 Nice, France (e-mail: cedric.richard@unice.fr).}}

%
%

\markboth{IEEE SIGNAL PROCESSING LETTERS,~VOL.~xx,~2022}%
{GAO \MakeLowercase{\textit{et al.}}: Adaptive Random Fourier Features Gaussian Kernel LMS Algorithm}
%

\maketitle

\begin{abstract}
We propose the adaptive random Fourier features Gaussian kernel LMS (ARFF-GKLMS). Like most kernel adaptive filters based on stochastic gradient descent, this algorithm uses a preset number of random Fourier features to save computation cost. However, as an extra flexibility, it can adapt the inherent kernel bandwidth in the random Fourier features in an online manner. This adaptation mechanism allows to alleviate the problem of selecting the kernel bandwidth beforehand for the benefit of an improved tracking in non-stationary circumstances. Simulation results confirm that the proposed algorithm achieves a performance improvement in terms of convergence rate, error at steady-state and tracking ability over other kernel adaptive filters with preset kernel bandwidth.
\end{abstract}

\begin{IEEEkeywords}
Kernel LMS, random Fourier features, Gaussian kernel, stochastic gradient descent.
\end{IEEEkeywords}

%

\section{Introduction}
\label{sec:Intro}

\IEEEPARstart{T}{he} kernel least-mean-square (KLMS) algorithm was first introduced in~\cite{Liu2008b} by reformulating the LMS algorithm in reproducing kernel Hilbert spaces (RKHS). Since the KLMS can be easily implemented and has good tracking performance, it has become central in the family of kernel adaptive filters. The Gaussian kernel is commonly used with kernel adaptive filters because it has universal modeling capability, desirable smoothness, and numerical stability~\cite{Keerthi2003, Liu2010}. In particular, the Gaussian kernel LMS (GKLMS) has attracted substantial research interests as well as its variants~\cite{Richard2009, Chen2012a,gao2013kernel,Zhao2015, Wang2016, Gao2017}, and its theoretical performance has been extensively analyzed~\cite{Parreira2012, Chen2014,Gao2014, Gao2019, Gao2020}. However, the selection of an appropriate bandwidth for the Gaussian kernel to ensure good performance still remains a problem with GKLMS-type algorithms in practical use, especially for non-stationary environments.

Multi-kernel LMS (MKLMS) algorithms use a collection of kernels with predefined bandwidths. They were developed to alleviate the issue of kernel bandwidth selection, unfortunately at the cost of an extra computational overhead~\cite{Yukawa2012, Tobar2014}. The GKLMS algorithm with adaptive kernel bandwidth, in parametric vector-valued form, and in non-parametric functional form, was introduced independently in~\cite{Fan2016,Chen2016}, but without considering any non-negative constraint for the kernel bandwidth. On the other hand, the random Fourier features GKLMS (RFF-GKLMS) algorithm was proposed to make the GKLMS algorithms computationally more efficient with low performance penalty~\cite{Bouboulis2016}. Random Fourier features (RFF) were also considered in~\cite{Bouboulis2018} for distributed learning over networks and graphs with kernel adaptive filters in order to address nonlinear regression and classification tasks. The RFF principle was used with the kernel conjugate gradient algorithm~\cite{Xiong2019}. The Cauchy-loss conjugate gradient method based on multiple RFF was proposed in~\cite{Zhang2021} to improve robustness and reduce computational cost in the presence of non-Gaussian noises. Recently, several RFF kernel regression algorithms over graphs were proposed in~\cite{Elias2022}, and their conditions for convergence in the mean and mean-square sense were also studied.

To the best of our knowledge, no RFF-based algorithm has been proposed yet to adapt random Fourier features. In this letter, we overcome this lack by devising the adaptive random Fourier features GKLMS (ARFF-GKLMS) algorithm. Based on stochastic gradient descent, it updates the vectors and phase factors of the RFF in an online manner. The ARFF-GKLMS algorithm outperforms the RFF-GKLMS and the GKLMS in terms of convergence rate, steady-state error and tracking ability. More importantly, the proposed simple but effective principle of adaptive RFF can be readily incorporated into all existing RFF filtering algorithms to enhance their performance.

\section{Gaussian kernel-based methods with RFF} 
\label{sec:ApproxGaussian}

Consider an unknown system with input-output relation characterized by the following nonlinear model:
\begin{equation}
    \label{eq:yn0}
    y_n = f^\star(\vx_n) + z_n
\end{equation}
where $f^\star(\cdot)$ is an unknown function to be identified in a given RKHS $\HH$ endowed with a kernel $\kappa(\cdot,\cdot)$, and $\vx_n\in\R^L$ is the original input data. The nonlinear desired output $y_n\in\R$ is corrupted by a zero-mean white Gaussian observation noise~$z_n$. Given input and noisy output data pairs $\{(\vx_n, y_n)\}_{n=1}^N$, we consider the following functional optimization problem:
\begin{equation}
    \label{eq:CostFun1}
   \min_{f \in \HH} \,\, \sum_{n=1}^N \ell \big(f(\vx_n), y_n \big) + \frac{\lambda}{2} \|f\|^2_\HH
\end{equation}
where $\ell(\cdot, \cdot)$ denotes any convex loss function, $\lambda>0$ is the regularization parameter, and $\|\cdot\|_\HH$ is the RKHS norm induced by its inner product. The Representer Theorem ensures that the solution $f(\cdot)$ can be represented by a kernel expansion in terms of training data~\cite{Sch00}:
\begin{equation}
    \label{eq:fn1}
    f(\cdot) = \sum_{n=1}^N w_n \kappa(\cdot, \vx_n) = \vw_N^\top \vkappa_N(\cdot)
\end{equation}
with $\vw_N=[w_1,\ldots,w_N]^\top$ the coefficient vector to determine and $\vkappa_N(\cdot)=\big[\kappa(\cdot, \vx_1), \ldots, \kappa(\cdot, \vx_N)\big]^\top$ the kernelized input.

Instead of using the kernel trick~\cite{Schlkopf2001, Herbrich2001}, which implicitly maps the data into a feature space, the input data can be explicitly mapped to a finite low-dimensional Euclidean space by a random Fourier feature nonlinear map, $\vz: \R^L \to \R^D$. Hence, the kernel evaluation step can be approximated as follows~\cite{Rahimi2007}:
\begin{equation}
    \kappa(\vx, \vx') = \langle \varphi(\vx), \varphi(\vx') \rangle \approx \vz(\vx)^\top \vz(\vx').
\end{equation}
A continuous and shift-invariant kernel $\kappa(\vx, \vx') =\kappa(\vdelta)$ with $\vdelta=\vx-\vx'$ defined on $\R^L$ is positive definite if, and only if, $\kappa(\vdelta)$ is the Fourier transform of a non-negative measure~\cite{Rudin2017}. When the kernel $\kappa(\vdelta)$ is properly scaled, Bochner's theorem guarantees that its Fourier transform:
\begin{equation}
    \label{eq:PDF}
    p(\vomega) =  \frac{1}{(2\pi)^L} \int_{\R^L}\kappa(\vdelta) \exp{(-j\vomega^\top\vdelta)}d\vdelta
\end{equation}
is a proper probability distribution~\cite{Rahimi2007}, where $j=\sqrt{-1}$. Defining $\zeta_\vomega(\vx)=\exp(j\vomega^\top \vx)$, we obtain:
\begin{equation}
    \label{eq:approxPDF}
    \begin{split}
    \kappa(\vx- \vx') &= \int_{\R^L} p(\vomega) \exp{\left(j \vomega^\top(\vx-\vx')\right)} d\vomega \\
                             &= \E_\vomega\big[\zeta_\vomega(\vx)^{\rm H} \zeta_\vomega(\vx')\big]
    \end{split}
\end{equation}
where $(\cdot)^{\rm H}$ denotes the Hermitian transpose operator. When~$\vomega$ is drawn from $p(\vomega)$, $\zeta_\vomega(\vx)^{\rm H} \zeta_\vomega(\vx')$ then provides an unbiased estimate of $\kappa(\vx, \vx')$. Because $\kappa(\vx, \vx')$ is real-valued, replacing $\exp{\left(j \vomega^\top(\vx-\vx')\right)}$ by its real part $\cos\left(\vomega^\top(\vx - \vx')\right)$ leads to a real-valued random feature for kernel $\kappa$. By defining mapping $z_{\vomega, b}(\vx)=\sqrt{2}\cos(\vomega^\top \vx + b)$, then the real-valued kernel function can be expressed as~\cite{Rahimi2007}:
\begin{equation}
    \label{eq:Kappx}
    \kappa(\vx, \vx') =  \E_{\vomega, b}\big[z_{\vomega, b}(\vx)^\top z_{\vomega, b}(\vx')\big]
\end{equation}
where $b$ is drawn from the uniform distribution on $[0, 2\pi]$. Using~\eqref{eq:Kappx} with the Gaussian kernel mentioned before, the latter can be approximated by $D$ random Fourier features and random phase factors:
\begin{equation}
    \label{eq:GaussianApprox}
    \begin{split}
    \kappa(\vx, \vx') &= \exp\left( -\|\vx - \vx'\|^2/2\xi^2\right) \\
                            & \approx \frac{1}{D} \sum_{m=1}^D z_{\vomega_m, b_m}(\vx) z_{\vomega_m, b_m}(\vx')
    \end{split}
\end{equation}
with $\xi>0$ the kernel bandwidth. Each $\vomega_m$ is obtained by sampling $p(\vomega) = \big(\xi/\sqrt{2\pi}\big)^D \exp\big(-\xi^2\|\vomega\|^2/2\big)$ beforehand, that is, $\vomega \!\sim\! \N(\bs 0_D, \xi^{-2}\mI_D)$. On the other hand, each $b_m$ is obtained by sampling $\U([0, 2\pi])$~\cite{Bouboulis2016, Bouboulis2018}. A high order $D$ can improve the approximation of~\eqref{eq:GaussianApprox}. However, a trade-off between accuracy and complexity has to be reached. Assume that the feature map $\vz_\Omega: \R^L \! \to \! \R^D$ is defined as:
\begin{equation}
    \label{eq:zomega}
    \vz_\Omega(\vx) \!=\! \sqrt{2/D} \big[ \cos(\vomega_1^\top \vx + b_1), \ldots, \cos(\vomega_D^\top \vx + b_D) \big]^\top.
\end{equation}
Substituting~\eqref{eq:zomega} into~\eqref{eq:GaussianApprox}, the kernelized input vector in~\eqref{eq:fn1} can be approximated by:
\begin{equation}
    \label{eq:ApproxKernel}
    \vkappa_N(\cdot) \approx \big[ \vz_\Omega(\vx_1)^\top \vz_\Omega(\cdot), \ldots, \vz_\Omega(\vx_N)^\top \vz_\Omega(\cdot) \big].
\end{equation}
By using approximation~\eqref{eq:ApproxKernel} and a sufficiently large order $D$, function~\eqref{eq:fn1} can be reformulated as:
\begin{equation}
    \label{eq:Estimate-2}
    f(\cdot) \approx \vw_N^\top \big[\vz_\Omega(\vx_1), \ldots, \vz_\Omega(\vx_N)\big]^\top\vz_\Omega(\cdot)
\end{equation}
We shall rewrite~\eqref{eq:fn1} as follows:
\begin{equation}
    \label{eq:Estimate}
    f(\cdot) = \valpha^\top\vz_\Omega(\cdot)
\end{equation}
with the modified $(D\times 1)$-dimensional weight vector $\valpha$ and the RFF nonlinear transformation $\vz_\Omega(\cdot)$:
\begin{equation}
    \label{eq:valpha}
    \valpha = \big[\vz_\Omega(\vx_1), \ldots, \vz_\Omega(\vx_N)\big] \, \vw_N
\end{equation}
Based on model~\eqref{eq:Estimate}, we can now derive a linear adaptive filtering strategy based on the LMS for updating $\valpha$ based on the $(D\times 1)$-dimensional RFF representation $\vz_\Omega(\cdot)$ of data.

\section{Adaptive random Fourier features GKLMS}
\label{sec:ARFF-GKLMS}
In this section, we introduce the proposed ARFF-GKLMS algorithm. Model~\eqref{eq:Estimate} shows that, although it no longer required to evaluate Gaussian kernel functions, the preset kernel bandwidth $\xi$ still plays a prominent role through the Gaussian vectors $\vomega_m$ sampled from $\N(\bs 0_D, \xi^{-2}\mI_D)$. Note that a parallel can be drawn between these vectors $\vomega_m$ and the dictionary elements usually considered with KAF algorithms; see, e.g.,~\cite{Liu2008b,Richard2009,Engel2004}. We shall now consider adjusting vectors $\{\vomega_m\}_{m=1}^D$ to improve the performance of RFF-based algorithms.

Consider the mean-square error cost function defined by:
\begin{equation}
    \label{eq:CostFun2}
    \L(\valpha, \vomega_\Omega, \vb_\Omega) = \E\left\{ \left|y_n - \valpha^\top \vz_\Omega(\vx_n)\right|^2 \right\}
\end{equation}
with $\vomega_\Omega=(\vomega_1, \cdots, \vomega_D)$ and $\vb_\Omega=(b_1, \cdots, b_D)$. We aim to estimate these optimal variables $\valpha$, $\vomega_\Omega$, and $\vb_\Omega$ by solving the following optimization problem of identifying the nonlinear system described by model~\eqref{eq:Estimate}:
\begin{equation}
    \label{eq:CostFun3}
    \min_{\valpha, \vomega_\Omega, \vb_\Omega} \L(\valpha, \vomega_\Omega, \vb_\Omega).
\end{equation}
Following to the steepest-descent principle, the weight vector $\valpha_{n+1}$ at time $n+1$ can be evaluated by updating the weight vector $\valpha_{n}$ at time $n$ as follows:
\begin{equation}
    \label{eq:valpha0}
    \valpha_{n+1} = \valpha_n + \frac{1}{2} \eta_\alpha \left[-\frac{\partial{\L(\valpha, \vomega_\Omega, \vb_\Omega)}}{\partial{\valpha}}\right]
\end{equation}
where $\eta_\alpha>0$ denotes the learning step-size. The gradient vector of $\L(\valpha, \vomega_\Omega, \vb_\Omega)$ with respect to $\valpha$ is approximated by its instantaneous value, i.e.,
\begin{equation}
    \label{eq:valphaGradient}
    \frac{\partial{\L(\valpha, \vomega_\Omega, \vb_\Omega)}}{\partial{\valpha}} \approx -2 e_n \vz_{\Omega,n}(\vx_n).
\end{equation}
with $e_n = y_n - \valpha_n^\top \vz_{\Omega,n}(\vx_n)$ the instantaneous estimation error. Substituting the stochastic subgradient~\eqref{eq:valphaGradient} into~\eqref{eq:valpha0}, we arrive at the update relation of the ARFF-GKLMS algorithm:
\begin{equation}
    \label{eq:valpha1}
    \valpha_{n+1} = \valpha_n + \eta_\alpha e_n \vz_{\Omega,n}(\vx_n)
\end{equation}
with $\valpha_n=[\alpha_{1,n}, \ldots, \alpha_{D,n}]^\top$ the weight vector, and $\vz_{\Omega,n}(\vx_n)$ the adaptive random Fourier features transformation vector:
\begin{equation}
    \label{eq:vh}
    \vz_{\Omega,n}(\vx_n) = \big[ \cos(\vomega_{1,n}^\top \vx_n + b_{1,n}), \ldots, \cos(\vomega_{D,n}^\top \vx_n + b_{D,n}) \big]^\top.
\end{equation}
Now we apply the steepest-descent principle to~\eqref{eq:CostFun2} in order to update the $m$-th vector $\vomega_{m,n}$:
\begin{equation}
    \label{eq:vomega0}
    \vomega_{m,n+1} = \vomega_{m,n} + \frac{1}{2} \eta_\omega \left[-\frac{\partial{\L(\valpha_n, \vomega_\Omega, \vb_\Omega)}}{\partial{\vomega_m}}\right]
\end{equation}
for $m=1,\ldots,D$, where $\eta_\omega>0$ is the corresponding learning step-size. Applying the chain rule to take the partial derivative of~\eqref{eq:CostFun2} with respect to $\vomega_m$, we obtain:
\begin{equation}
    \label{eq:omegaGradient}
    \begin{split}
    \frac{\partial{\L(\valpha, \vomega_\Omega, \vb_\Omega)}}{\partial \vomega_m} &= \frac{\partial{\L(\valpha, \vomega_\Omega, \vb_\Omega)}}{\partial z_{\Omega,m,n}(\vx_n)} \cdot \frac{\partial{z_{\Omega,m,n}(\vx_n)}}{\partial \vomega_m} \\
    &\approx 2e_n \alpha_{m,n} \sin(\vomega_{m,n}^\top \vx_n + b_{m,n}) \vx_n
    \end{split}
\end{equation}
where the subgradient vector $\partial{\L(\valpha, \vomega_\Omega, \vb_\Omega)}/\partial \vomega_m$ is replaced by its instantaneous estimate, i.e., the stochastic subgradient. Substituting~\eqref{eq:omegaGradient} into~\eqref{eq:vomega0}, the update equation is given by:
\begin{equation}
    \label{eq:vomega2}
    \vomega_{m,n+1} = \vomega_{m,n} - \eta_\omega e_n \alpha_{m,n} \sin(\vomega_{m,n}^\top \vx_n + b_{m,n}) \vx_n.
\end{equation}
Likewise, we can obtain the recursive relation of the $m$-th phase factor $b_{m,n}$:
\begin{equation}
    \label{eq:b2}
    b_{m,n+1} = b_{m,n} - \eta_b e_n \alpha_{m,n} \sin(\vomega_{m,n}^\top \vx_n + b_{m,n})
\end{equation}
for $m=1,\ldots,D$, with the learning step-size $\eta_b>0$. The procedures of the ARFF-GKLMS are listed in Algorithm~\ref{alg:Framwork}.

Before going further, two important points need to be given attention. First, problem~\eqref{eq:CostFun3} is no longer convex with respect to variables $(\valpha,\vomega_\Omega, \vb_\Omega)$. We shall however observe in the next section that, thanks to the adaptation steps~\eqref{eq:vomega2} and~\eqref{eq:b2}, the ARFF-GKLMS algorithm offers a fast convergence rate, low steady-state error, and good tracking ability, in particular when processing with non-stationary systems. More importantly, due to its simplicity, the VRFF method can be readily applied to other RFF-based algorithms.
Secondly, the adaptation steps~\eqref{eq:vomega2} and~\eqref{eq:b2} no longer guaranty that the $\{\vomega_{m,n}\}$ are driven by any Gaussian distribution $\N(\bs 0_D, \xi^{-2}\mI_D)$ as the algorithm progresses. This does not allow us to establish a correspondence between the $\{\vomega_{m,n}\}$ and the bandwidth $\xi_n$ of a Gaussian kernel. Further work will be carried out to give a better insight in understanding the properties of the nonlinear mapping.

\begin{algorithm}[!htbp]
\caption{ARFF-GKLMS algorithm}
\label{alg:Framwork}
\begin{algorithmic}[1]

\STATE \textbf{Initialization:} \\
\STATE Set the step-sizes $\eta_\alpha$, $\eta_\omega$, $\eta_b$, and the kernel bandwidth $\xi$. \\
\STATE Generate random $\vomega_{m,1}$ and $b_{m,1}$ for $m=1, 2, \ldots, D$.
\STATE \textbf{Input:} $\big\{(\vx_n, y_n)\big\}$, $n=1,2,\ldots, N.$ \\
\STATE \textbf{for} $n=1,2,\cdots, N$ \textbf{do}

\STATE
\begin{adjustwidth}{0.4cm}{0cm}
Update $\valpha_{n+1}$ via~\eqref{eq:valpha1}.
\end{adjustwidth}
\STATE
\begin{adjustwidth}{0.4cm}{0cm}
\textbf{for} $m=1, 2, \ldots, D$ \textbf{do}
\end{adjustwidth}
\STATE
\begin{adjustwidth}{0.8cm}{0cm}
Update $\vomega_{m,n+1}$ via~\eqref{eq:vomega2};
\end{adjustwidth}
\STATE
\begin{adjustwidth}{0.8cm}{0cm}
Update $b_{m,n+1}$ via~\eqref{eq:b2}.
\end{adjustwidth}
\STATE
\begin{adjustwidth}{0.4cm}{0cm}
\textbf{end for}
\end{adjustwidth}

\STATE \textbf{end for}

\STATE \textbf{Output:} $f(\vx_N)$.

\end{algorithmic}
\end{algorithm}

\section{Simulation Results}
\label{sec:Simulations}

In this section, we shall present two simulation examples to validate the improved performance of the ARFF-GKLMS algorithm compared to its RFF-GKLMS counterpart on the one hand, and to the classical GKLMS algorithm with coherence sparsification (CS) criterion~\cite{Richard2009} on the other hand. All the simulated curves were obtained by averaging over 200 independent Monte Carlo runs.

\subsection{Stationary Nonlinear System Identification}
\label{subsec:Example1}

Consider first the stationary nonlinear system defined by: 
\begin{equation*}
    \label{eq:yn1}
    y_n = \vkappa^\top_{\xi^\star}(\vx_n)\, \vw^\star + z_n
\end{equation*}
where $z_n$ denotes a zero-mean Gaussian observation noise at a $\text{SNR}=15\text{dB}$, and $\vw^\star$ the optimal weight vector is given by
\begin{equation*}
    \label{eq:w_opt}
    \vw^\star=[0.756, \,-1.384, \,-0.101, \,0.445, \,-0.565, \,0.134]^\top.
\end{equation*}
The kernelized input vector $\vkappa_{\xi^\star}(\vx_n)$ was constructed based on the Gaussian kernel with bandwidth $\xi^\star=0.95$ and the dictionary elements defined by:
\begin{equation*}
    \label{eq:D}
    \D = \left\{\! \left[\begin{matrix}0.17 \\ -1.92\end{matrix}\right]\!,  \left[\begin{matrix}-1.62 \\ -0.18\end{matrix}\right]\!,  \left[\begin{matrix}0.52 \\ 1.55\end{matrix}\right]\!,  \left[\begin{matrix}2.90 \\ 1.92\end{matrix}\right]\!,  \left[\begin{matrix}-2.01 \\ -2.47\end{matrix}\right]\!,  \left[\begin{matrix}2.66 \\ -0.82\end{matrix}\right] \!\right\}.
\end{equation*}
The input sequence $x_n = \rho x_{n-1} + \sqrt{1 - \rho^2} u_n$ was generated with correlation coefficient $\rho=0.5$ and $u_n$ a random sequence governed by the i.i.d. standard normal distribution. The input data vector was defined as $\vx_n=[x_n, x_{n-1}]^\top$. The step-size~$\eta_\alpha$ was set to $0.2$ for the GKLMS-CS, $0.01$ for the RFF-GKLMS, and $0.005$ for the ARFF-GKLMS, respectively. Both step-sizes~$\eta_\omega$ and $\eta_b$ were set to $1$. The number of RFF was set to $D=48$. The kernel bandwidth of the Gaussian kernel used by the GKLMS-CS was set to $\xi=0.95$, and the threshold of the CS criterion $\delta_\kappa$ was set to $0.7$ to obtain the final dictionary size $M=51$ for the comparisons of learning curves of transient EMSE. 

\begin{figure}[!htbp]
	\centering
    \subfigure[] 
	{\includegraphics[trim = 4mm 1mm 13mm 6.5mm, clip, width=0.25\textwidth]{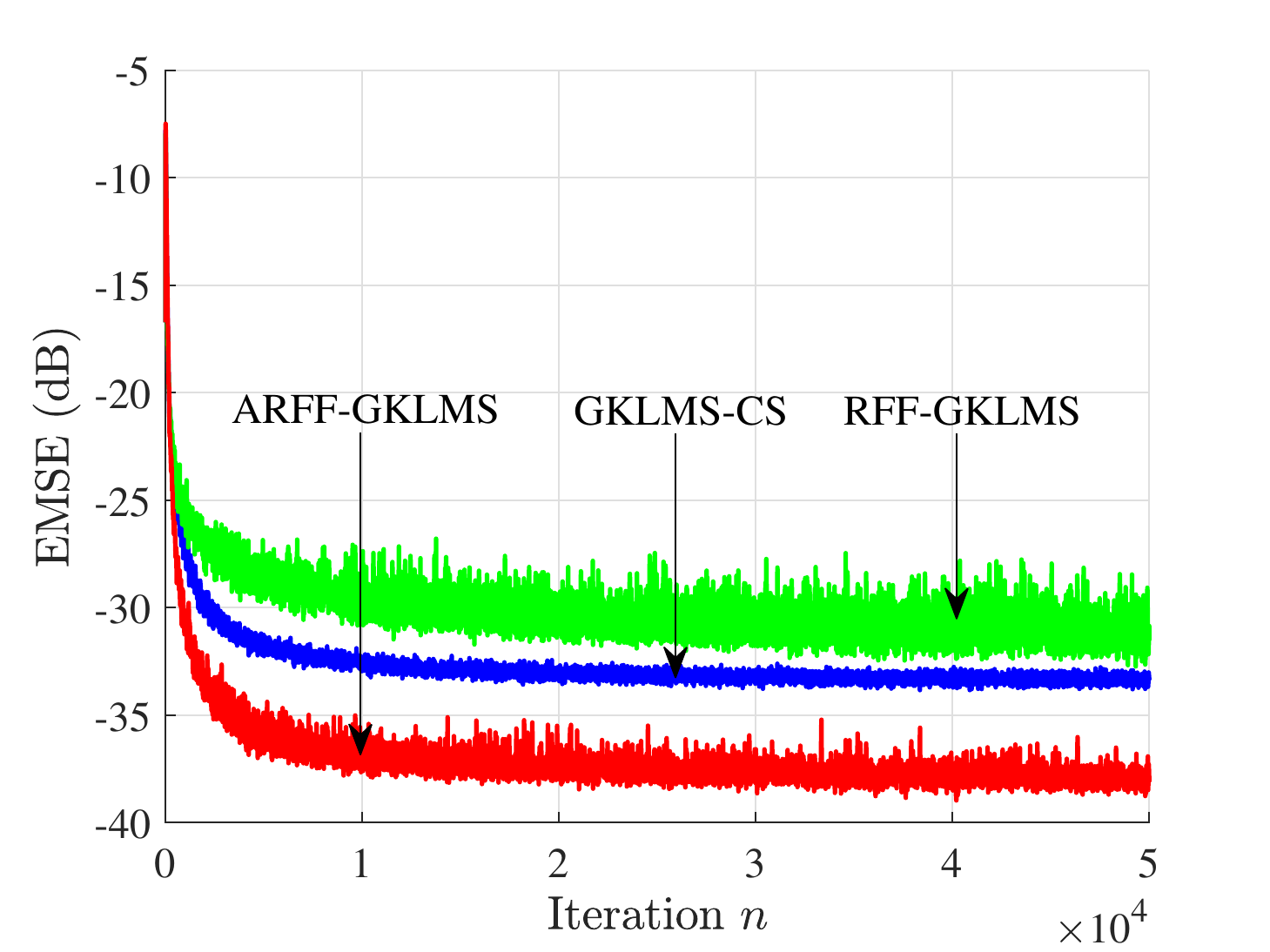}}
    \subfigure[] 
	{\includegraphics[trim = 8mm 0mm 13mm 6.5mm, clip, width=0.24\textwidth]{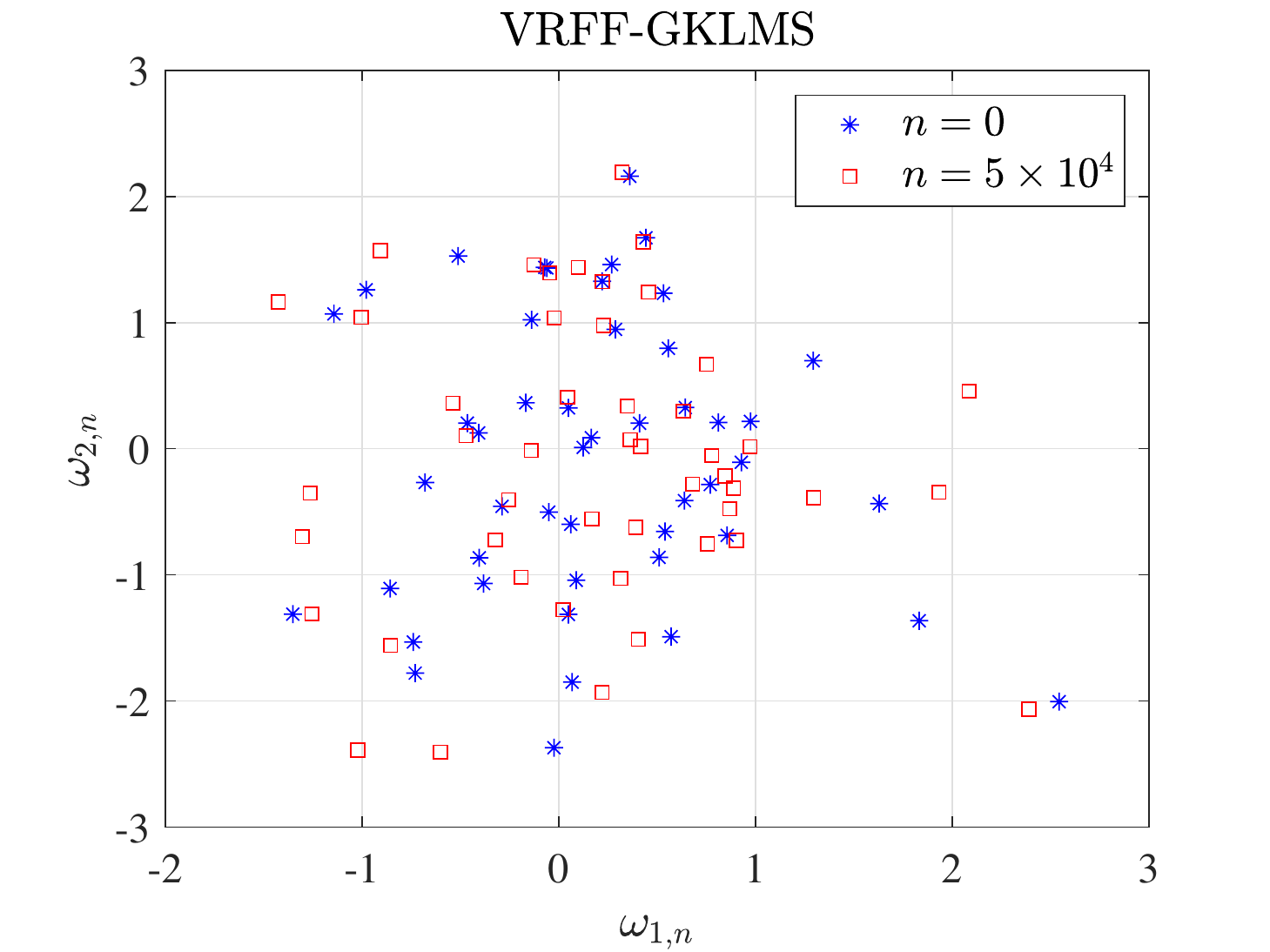}} \\
    \subfigure[] 
	{\includegraphics[trim = 0mm 0mm 13mm 6.5mm, clip, width=0.24\textwidth]{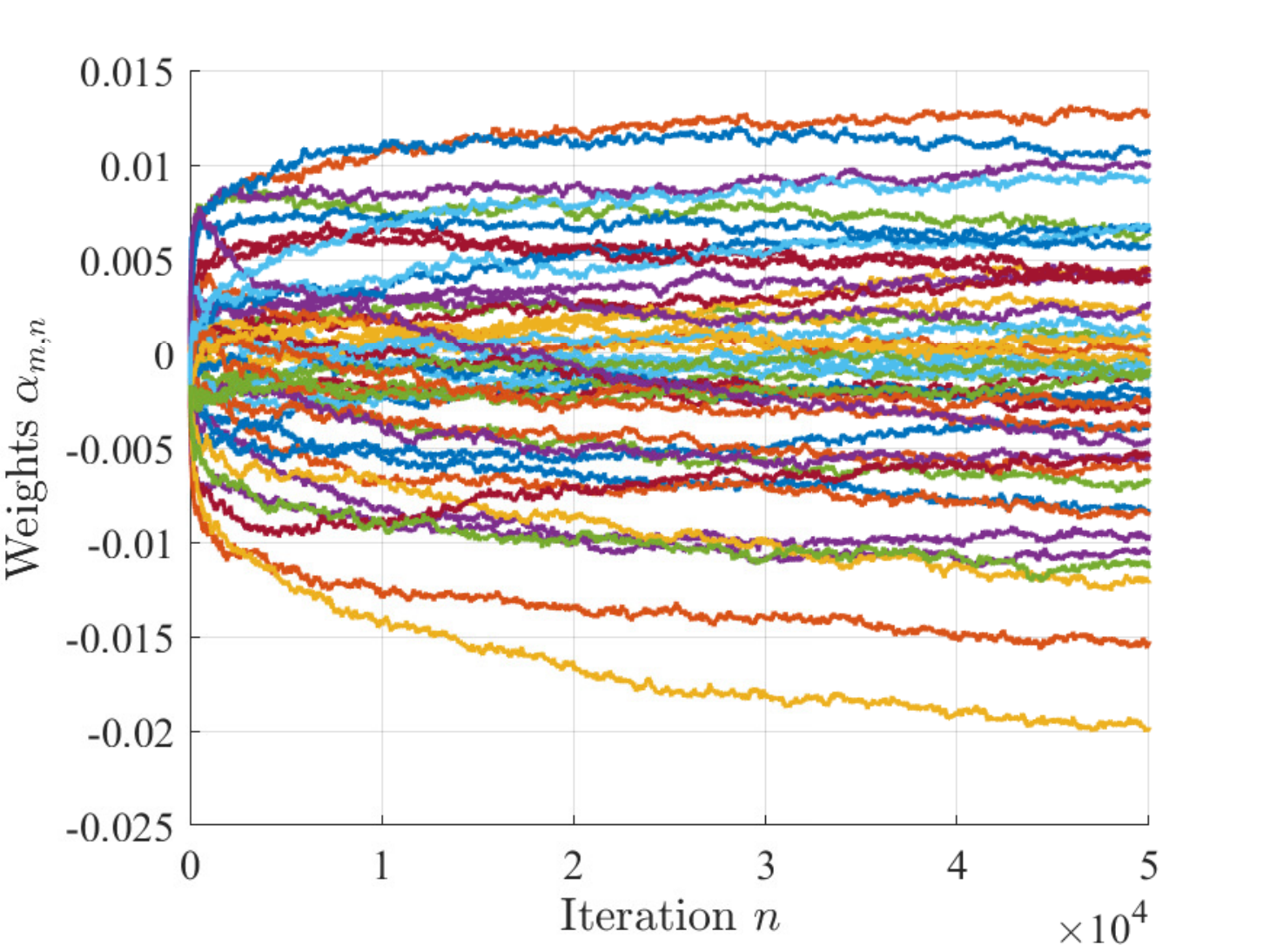}}
    \subfigure[] 
	{\includegraphics[trim = 4mm 0mm 13mm 2mm, clip, width=0.24\textwidth]{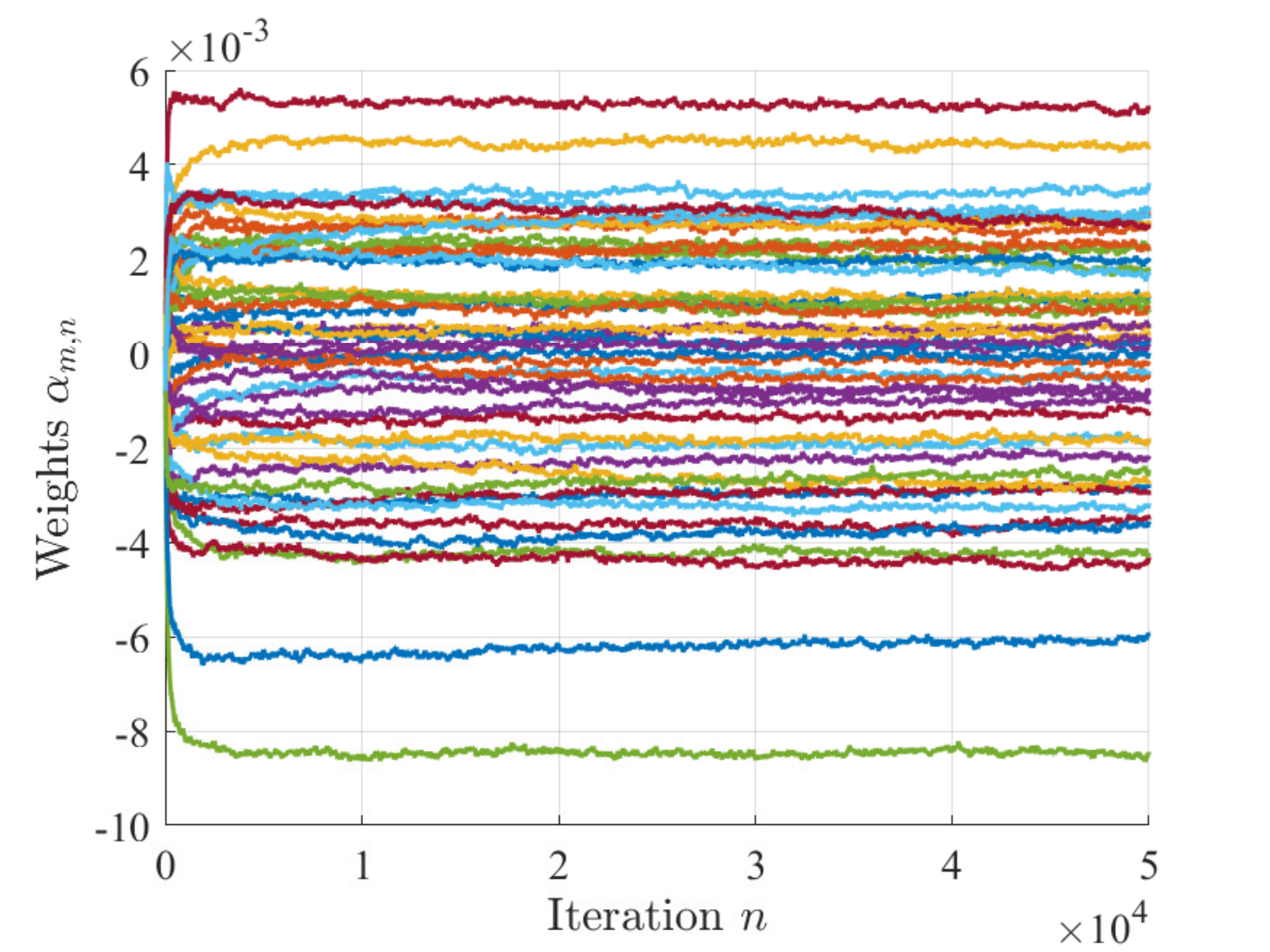}} \\
    \subfigure[] 
	{\includegraphics[trim = 4mm 0mm 13mm 6.5mm, clip, width=0.24\textwidth]{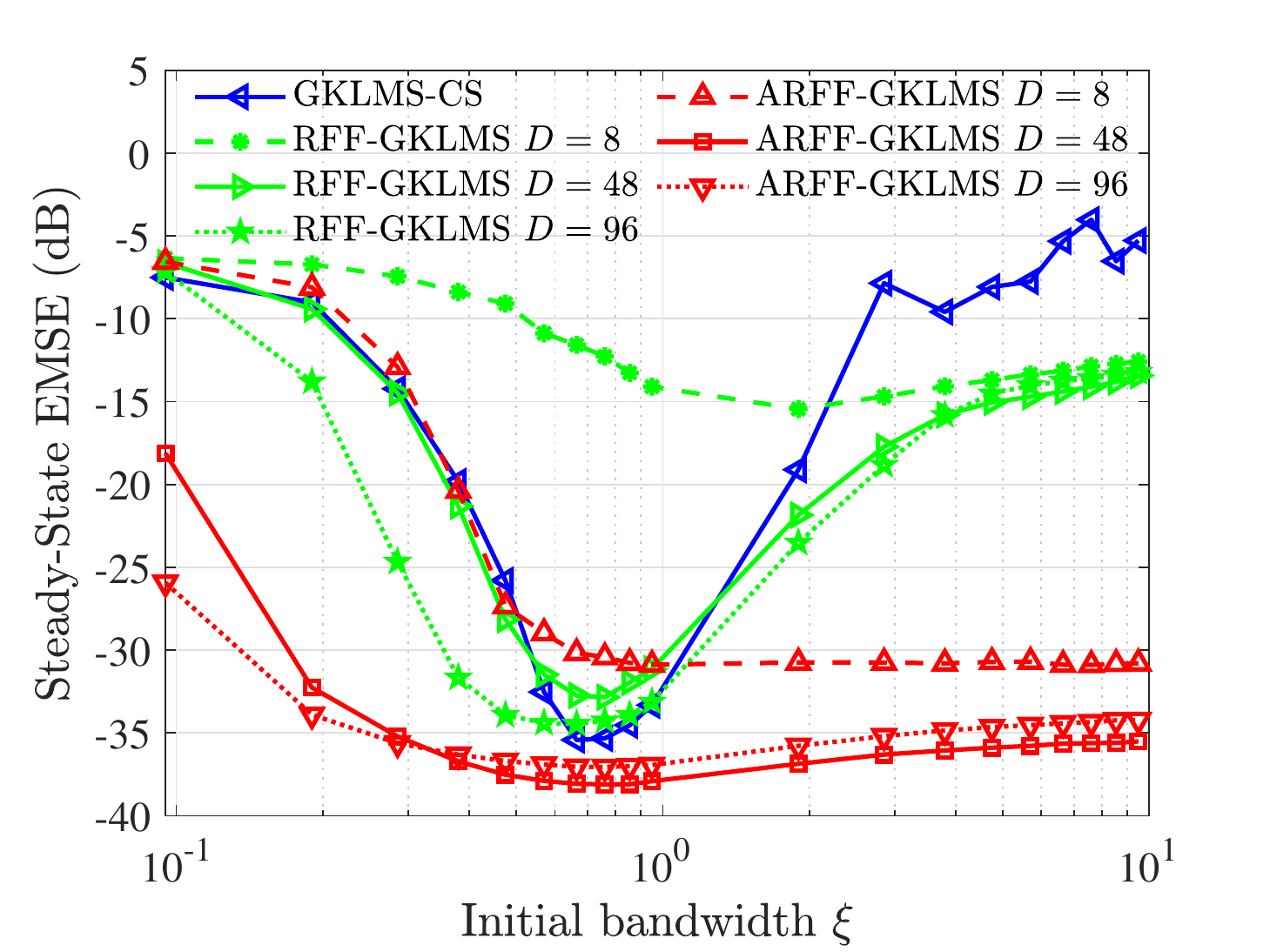}}
    \subfigure[] 
	{\includegraphics[trim = 3mm 0mm 13mm 8mm, clip, width=0.24\textwidth]{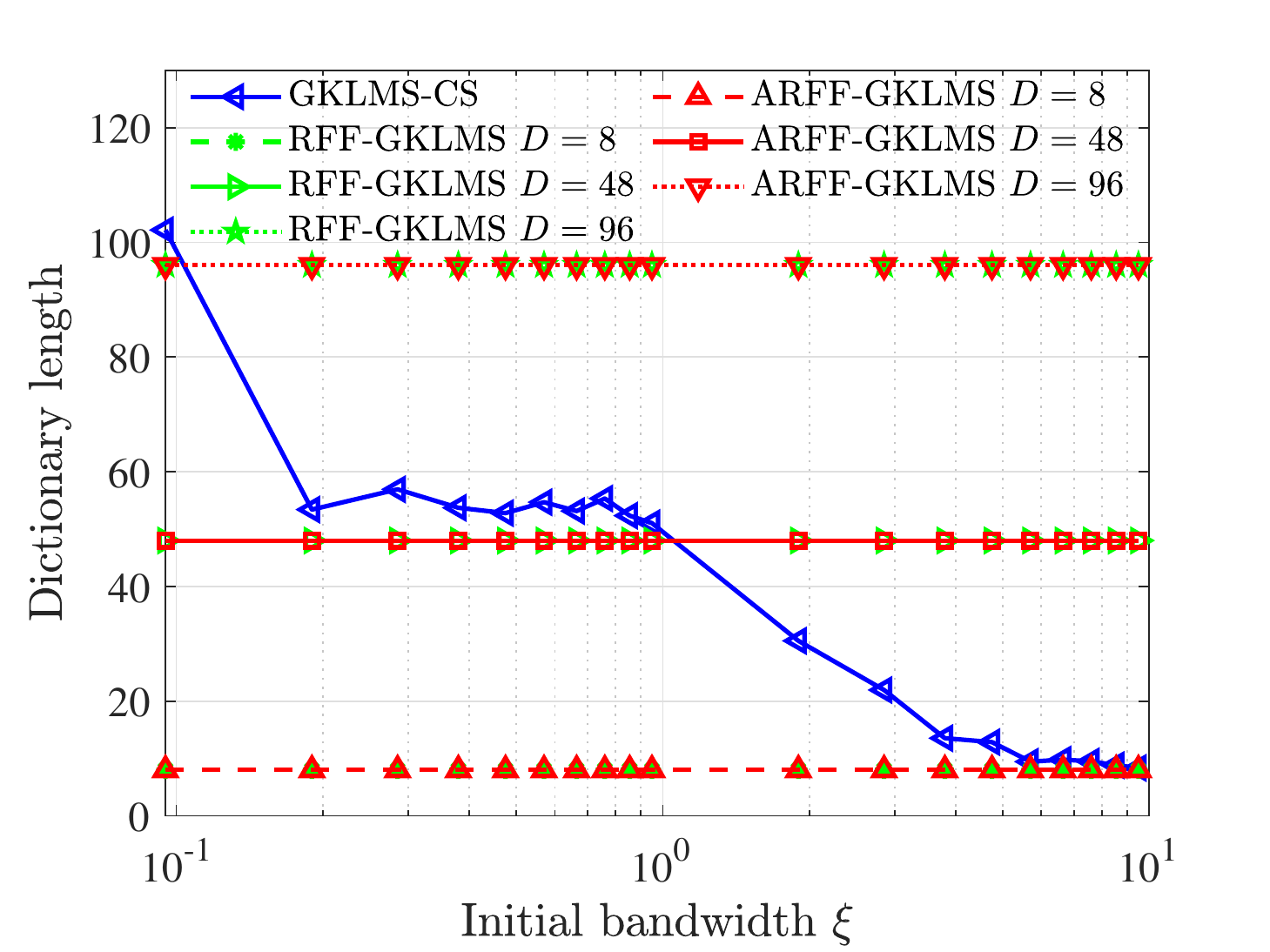}}
    \caption{Simulation results for stationary nonlinear system identification. (a) Learning curves of transient EMSE for $\xi=0.95$, $M=51$, and $D=48$. (b) Location of vectors $\vomega_n$ $(D=48)$. (c) Weights evolution for the RFF-GKLMS $(D=48)$. (d) Weights evolution for the ARFF-GKLMS $(D=48)$. (e) Steady-state EMSE versus $\xi$ initial setting. (f) Dictionary length versus different initial $\xi$.}
	\label{fig:ExampI}
\end{figure}

Fig.~\ref{fig:ExampI}(a) shows that ARFF-GKLMS algorithm significantly outperformed the RFF-GKLMS and the GKLMS-CS algorithms in terms of convergence rate and steady-state excess-mean-square error (EMSE), which is the mean of the last $5\times10^3$ entries of the ensemble-average learning curve of EMSE. Correspondingly, Figs.~\ref{fig:ExampI}(c) and \ref{fig:ExampI}(d) show that the weight coefficients of the ARFF-GKLMS converged faster than those of the RFF-GKLMS. Fig.~\ref{fig:ExampI}(b) shows the location of vectors $\vomega_n$ at the beginning and at the end of the optimization process. Fig.~\ref{fig:ExampI}(e) shows that $\xi$ setting has a strong effect on the performance of the GKLMS-CS algorithm, which thus needs to be carefully initialized based on side information or preliminary tests. After a transient stage, the dictionary size reaches a maximum value $M$ determined by the CS criterion. Fig.~\ref{fig:ExampI}(f) shows that $M$ gradually decreases as the kernel bandwidth $\xi$ increases. We observe on Fig.~\ref{fig:ExampI}(e) that the EMSE at steady state of the RFF-GKLMS algorithm is sensitive to large orders $D$ and to kernel bandwidth $\xi$ setting. In contrast, we can notice that the EMSE at steady-state of the ARFF-GKLMS algorithm is robust with respect to kernel bandwidth~$\xi$ initialization, particularly for large $\xi$. Nevertheless, the algorithm suffers from performance degradation when both $D$ and the kernel bandwidth are small, as a result of the poor approximation capacities of the kernel model in that case.

\subsection{Non-stationary Nonlinear System Identification}
\label{subsec:Example2}

Consider the following non-stationary nonlinear system with an abrupt change at time instant $n=1 \times 10^4$: 
\begin{equation*}
    \label{eq:Exam2}
    \begin{cases}
    d_n=& \hspace{-2.85mm} \big[0.8 - 0.5 \exp(-d^2_{n-1})\big]d_{n-1} + 0.1\sin(d_{n-1}\pi) \\
    & \hspace{-2.5mm} - \,\big[0.3 + 0.9 \exp(-d^2_{n-1})\big]d_{n-2}, \\
    & \hspace{+10mm} \text{for} \,\,\, 0\leq n \leq 5 \times 10^3 , \\
    d_n=& \hspace{-2.85mm} \big[0.2 - 0.7 \exp(-d^2_{n-1})\big]d_{n-1} +\, 0.2\sin(d_{n-1}\pi) \\
    & \hspace{-2.5mm} - \,\big[0.8 + 0.8 \exp(-d^2_{n-1})\big]d_{n-2}, \\
    & \hspace{+10mm} \text{for} \,\,\, 5 \times 10^3 < n \leq 1 \times 10^4, \\
     y_n=& \hspace{-2.5mm} d_n + z_n,
    \end{cases}
\end{equation*}
with $d_{-1}=d_{-2}=0.1$, and $z_n$ a zero-mean white Gaussian noise at $\text{SNR}=25\text{dB}$. The input data vector consists of the nonlinear delayed system outputs $\vx_n=[d_{n-1}, d_{n-2}]^\top$. The step-size $\eta_\alpha$ was set to $0.05$ for the GKLMS-CS, and to $0.005$ for the RFF-GKLMS and the ARFF-GKLMS, respectively. Both step-sizes $\eta_\omega$ and $\eta_b$ were set to $0.05$. The threshold of the CS criterion $\delta_\kappa$ and the order $D$ were set to $0.9$ and $96$, respectively. In order to test the ability of the ARFF-GKLMS to track nonstationary systems, the kernel bandwidth~$\xi$ was set to $0.3661$ for all algorithms.

\begin{figure}[!htbp]
	\centering
    \subfigure[] 
	{\includegraphics[trim = 4mm 1mm 8mm 6.5mm, clip, width=0.3\textwidth]{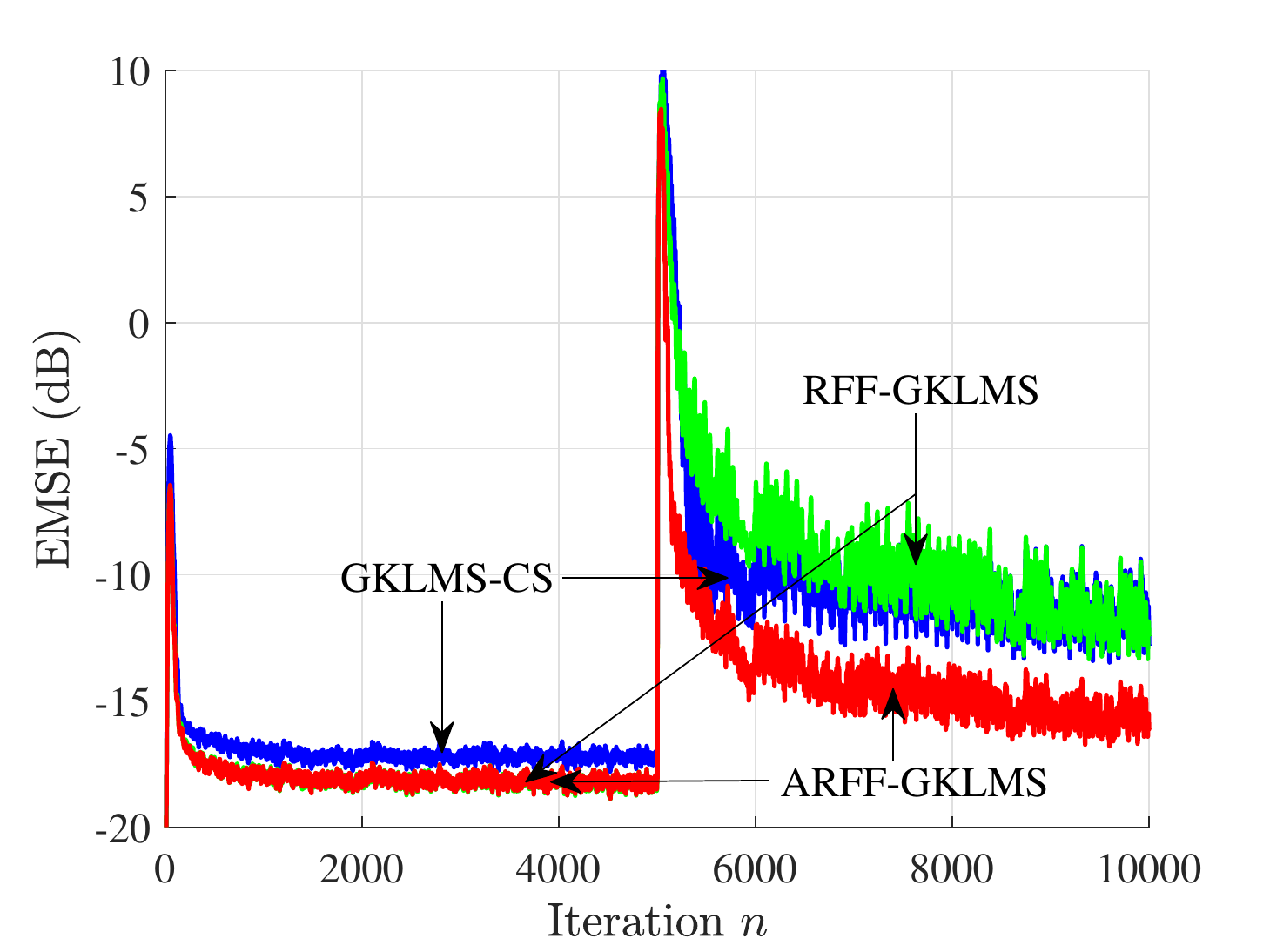}} 
    \subfigure[] 
	{\includegraphics[trim = 3mm 1mm 8mm 6.5mm, clip, width=0.25\textwidth]{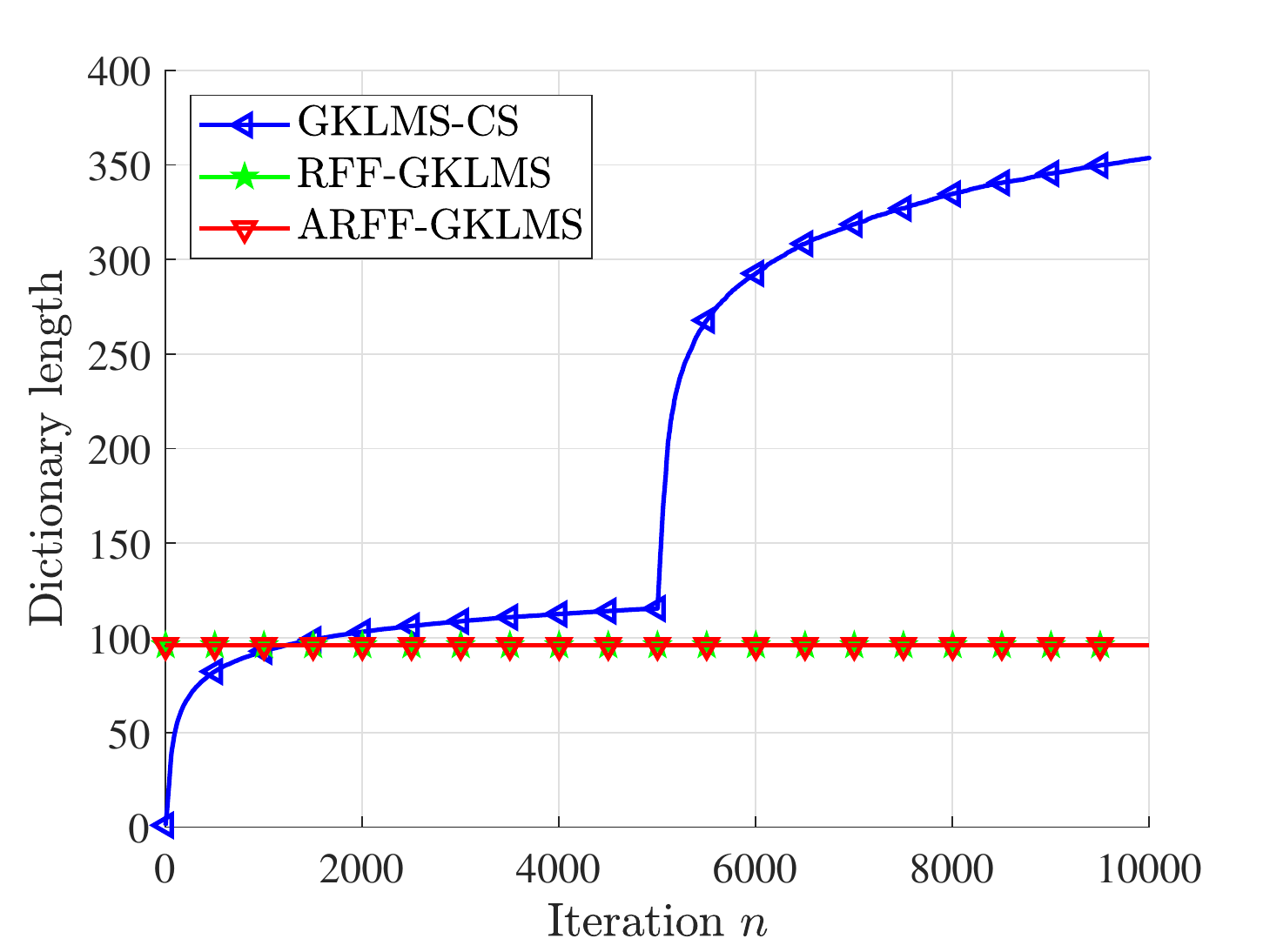}} 
    \subfigure[] 
	{\includegraphics[trim = 8mm 0mm 11mm 6.5mm, clip, width=0.23\textwidth]{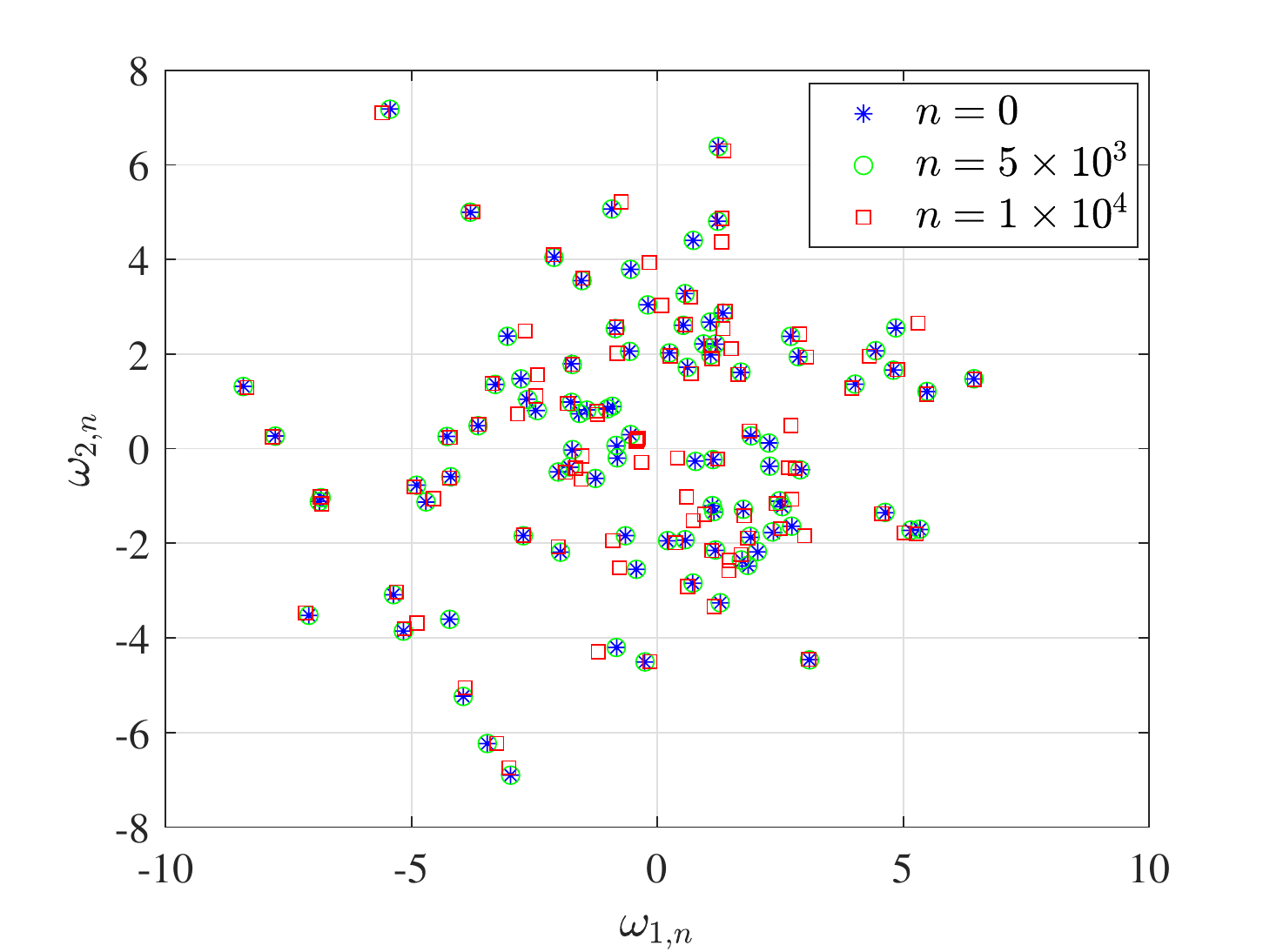}} 
    \caption{Simulation results for nonstationary nonlinear system identification. (a) Learning curves of transient EMSE . (b) Evolution of the dictionary length. (c) Variation in the locations of vectors $\vomega_n$ $(D=96)$.} 
	\label{fig:ExampII}
\end{figure}

Fig.~\ref{fig:ExampII}(a) shows that the ARFF-GKLMS has a good tracking ability, characterized by the lowest steady-state EMSE and the fastest convergence rate after the abrupt change, thanks to the online adaptation of~$\vomega_n$. Fig.~\ref{fig:ExampII}(b) shows that the dictionary length of the ARFF-GKLMS remains significantly smaller than that of the GKLMS-CS, which allows to save computation overhead. We can observe on Fig.~\ref{fig:ExampII}(c) that the locations of vectors $\vomega_n$ remained unchanged during the first stationary phase because the kernel bandwidth was carefully initialized for it, and then almost half of the $\vomega_n$ changed in order to adapt to the abrupt change.

\section{Conclusion}
\label{sec:Conclusion}

In this letter, we proposed a novel ARFF-GKLMS algorithm to adapt random Fourier features. This extra flexibility endows the algorithm with robustness and good tracking ability in non-stationary environments. The simulation results showed a significant performance improvement, both in transient and steady state. Since the step-sizes $\eta_\omega$ and $\eta_b$ have an important impact on the performance of the ARFF-GKLMS, variable step-size methods will be considered in future work. We will also apply a forward-backward splitting framework to eliminate the features with negligible contribution to the estimation performance. Finally, as mentioned before, further work will be carried out to give a better insight in understanding the properties of the nonlinear mapping resulting from the adaptation process.



\newpage


\balance

\bibliographystyle{IEEEtran}
\bibliography{ARFFGKLMS}

%
%

%

%
%
%




\end{document}